# A Framework for Transparent Reporting of Data Quality Analysis Across the Clinical Electronic Health Record Data Lifecycle

Melinda WASSELL [a,1], Kerryn BUTLER-HENDERSON [b], Karin VERSPOOR [a]
*[a] School of Computing Technologies, RMIT University, Melbourne, Australia*
[b] School of Nursing, Paramedicine and Healthcare Sciences, Charles Sturt University, NSW, Australia. [1] drmelinda.wassell@gmail.com
ORCiD ID: Melinda Wassell https://orcid.org/0000-0002-7139-3164

**Abstract.** Data quality (DQ) and transparency of secondary data are critical factors that delay the adoption of clinical AI models and affect clinician trust in them. Many DQ studies fail to clarify where, along the lifecycle, quality checks occur, leading to uncertainty about provenance and fitness for reuse. This study develops a framework for transparent reporting of DQ assessments across the clinical electronic health record (EHR) data lifecycle. The reporting framework was developed through iterative analysis to identify actors and phases of the clinical data lifecycle. The framework distinguishes between data-generating organizations and data-receiving organizations to allow users to map DQ parameters to stages across the data lifecycle. The framework defines 5 key lifecycle phases and multiple actors. When applied to the real-world dataset, the framework demonstrated applicability in revealing where DQ issues may originate. The framework provides a structured approach for reporting DQ assessments, which can enhance transparency regarding data fitness for reuse, supporting reliable clinical research, AI model development, and internal organisational governance. This work provides practical guidance for researchers to understand data provenance and for organisations to target DQ improvement efforts across the data lifecycle.

## 1. Introduction

Healthcare is experiencing increasing volumes of EHR data available for reuse, driven by advances in natural language processing and the increased adoption of structured data collection. This expansion creates more opportunities for data quality (DQ) issues to emerge at multiple points throughout the clinical EHR data lifecycle (1). Each new data source and interaction with data creates potential points of quality issues that may compromise the fitness for reuse. Without improved data quality reporting transparency, more data can simply lead to more quality issues.

The rapid adoption of artificial intelligence (AI) in healthcare is challenged by barriers related to data quality and trustworthiness with concerns about the reliability of models or research built on data whose provenance or quality is poorly understood (2).

Health data sharing is accelerating through interoperability and standardized data frameworks, such as the Australian Core Data for Interoperability (AUCDI)(3). When data is generated by multiple practitioners using multiple systems across a patient's healthcare team, it is necessary to have transparent quality assessment and reporting mechanisms that can trace data provenance and provide confidence in the data.

Additionally, healthcare organisations are increasingly leveraging their own clinical data for internal quality and safety improvement, clinical decision support, and to develop learning health systems (4, 5). As health care shifts towards data-driven care, new stakeholders such as administrators and funding bodies, who rely on accurate and timely information for patient care and organisational governance, are introduced.

Many studies that reuse EHRs for modelling address only a limited set of DQ parameters, such as completeness. DQ frameworks define what intrinsic, contextual, and technical parameters to assess (6), but not where along the lifecycle these assessments should occur. While lifecycle concepts are established (7, 8), existing frameworks do not provide structure for systematically categorizing and reporting where along the data lifecycle phase these occurred and by what actor. Most DQ analysis focuses on data at the point of reuse, after transformation for registries or for common data models (9), where data engineers may struggle to understand the factors that affected the initial data collection. Whilst the importance of DQ assessment at the point of data generation has been noted, organisational differences make this challenging (10).

Without clear provenance information and lifecycle-based quality reporting, researchers, AI developers and organisational stakeholders cannot adequately assess whether data is suitable for their intended uses or trace quality issues to their sources. Transparent reporting of where DQ assessments occur throughout the lifecycle is therefore essential to support data reuse and strengthen trust in secondary data use. Therefore, this paper will describe the development of a that creates transparency in DQ assessments along the clinical EHR data lifecycle.

## 2. Methods

We adapted and extended a model of the lifecycle of secondary use of clinical data (8) to develop a novel framework for reporting the areas along the clinical EHR data lifecycle where DQ needs to be assessed. An iterative process was used, involving clinicians and technical specialists to address stakeholder conflicts. Clinical EHR data from initial generation through to secondary reuse were assessed to document points at which data are created or modified.

Local knowledge experts started with two broad lifecycle phases: the data-generating organisation (DGO), where clinical data originates, and the data-receiving organisation (DRO), which is typically a registry or research organisation, and then further broke down lifecycle phases and defined actors at each phase.

To assist in systematic development and early application of the framework, an EHR dataset that has undergone thorough DQ analysis (11) was used to apply the model by ensuring lifecycle stages and actors were appropriate points to include. The EHR dataset originates from occupational health clinics in Australia and comprises 57,570 patient injury records. The organisation has many established governance processes for data use. The EHR data is mostly structured data capture, and the dataset has been established as relevant across the Australian workforce and occupational injuries (12). Once developed, the reporting framework was used to map potential DQ checks against the parameters defined in the expanded data quality assessment frameworks to ensure it was appropriate. Ethics approval; RMIT University #26603.

## 3. Results

The framework development process resulted in a structured model for transparent reporting of DQ assessments across the clinical EHR data lifecycle (Figure 1). The framework defines five key lifecycle phases and specifies the actors responsible at each phase.

**Figure 1.** Clinical EHR Data Lifecycle Phases for Transparent Reporting of Data Quality Analysis.

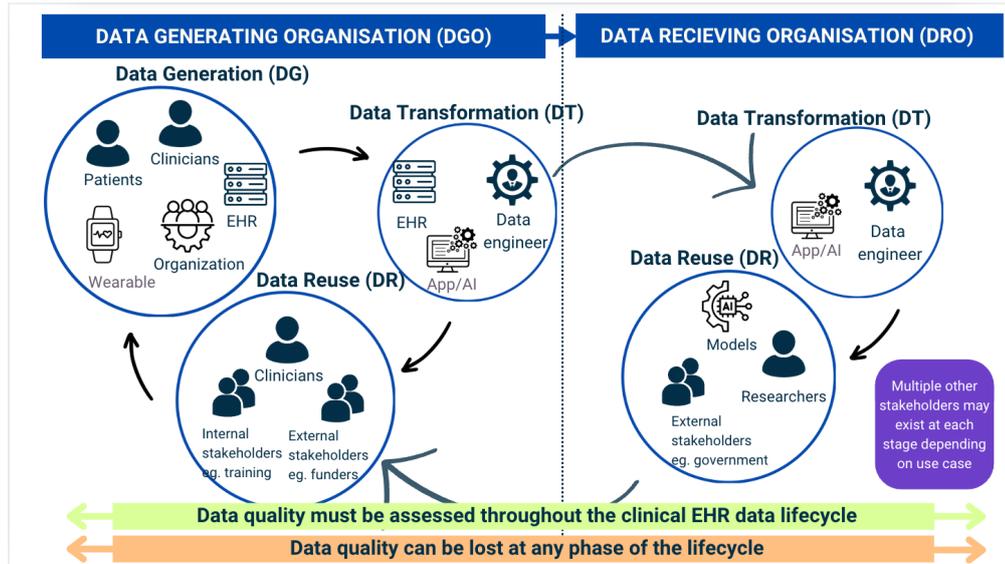

At the DGO, 3 lifecycle phases were included: data generation, data transformation, and data reuse. Our model describes actors at each phase relevant to the dataset used for validation. Patients, clinicians, and wearables can all directly generate data. The EHR system often generates data of its own, such as timestamps and calculations. Organization is a separate actor, as it often defines processes and policies governing clinicians' data entry, which can affect data quality or bias in entry. For example, some codes may be eligible for reimbursement, leading clinicians to code one of several similar codes to ensure applicability. EHR training procedures may also significantly impact DQ and are thus appropriate to label as such in DQ reporting.

As clinicians and healthcare organisations are responsible for data entered into records, wearables are considered actors within the DGO-DG phase. This distinction is critical, as many patient-used devices may not be regulatory-approved, posing a significant potential source of error. Highlighting whether the clinician took a clinical reading or it was generated by a software application automatically is critical for reusers of data to understand potential errors.

Data transformation (DT) refers to the processes of translating data into required formats or performing calculations. Actors are typically data engineers or automated by the EHR system itself.

Data reuse refers to the phase where DQ checks are conducted by the data reuser, such as by a clinician, researcher, or stakeholder. Stakeholders involved in data reuse include external funding bodies or government organisations. AI models built from EHR clinical data or clinical decision support tools also fit into this phase. As learning health systems evolve and data may undergo multiple iterations, defining the actors in the reuse phase becomes increasingly important.

Both DT and DR occur within the DGO and at the DRO because different use cases exist at each location, and the DRO often lacks visibility into organizational knowledge and contextual factors affecting data generation.

The datasets DQ checks were organized across intrinsic (completeness, conformance, and plausibility), contextual (timeliness, relevance, accessibility, and governance), and system/technical (operating platform and interoperability) DQ parameters in line with existing frameworks (6, 10).

The framework revealed that contextual parameters, such as governance and timeliness, were predominantly assessed at the DGO level, though different parameters may be relevant at the DGO versus the DRO level. The framework also demonstrated that similar DQ checks may need to be conducted at different lifecycle phases or for different actors, revealing quality issues that a single assessment would miss.

## 4. Discussion

The lifecycle reporting framework enables reusers to describe where DQ checks are conducted and which actors are involved in creating and modifying data. The framework supports transparency of data provenance and building trust in AI models by providing a standardized language for describing data's journey. Lifecycle-based reporting helps identify more specific areas of DQ issues and indicates whether interventions should target clinician training, organizational policies, or system design.

### 4.1. Relationship to Existing Data Quality Frameworks

The framework complements existing DQ frameworks and tools. The expanded and harmonised intrinsic data quality frameworks (6, 10) define what parameters to assess, whilst our framework specifies where assessments occur. Tools like Achilles (13) provide automated assessments, whilst our framework provides the semantic context for reporting of results. For example, the meaning of an Achilles report showing that 15% of diagnosis codes are missing improves when reported as: "DGO-DG-Organization (Policy: states Diagnosis only required only for billable encounter)" and "DRO-DT-Engineer (Mapping: 92% success)", revealing both policy and technical contributions to missingness.

The framework aligns with emerging requirements for data provenance documentation (14), such as the FAIR (Findable, Accessible, Interoperable, Reusable) principles (15). The lifecycle terminology can be incorporated into metadata documentation, to enable machine-readable provenance tracking. Additionally, where reuse occurs internally, the identification of organizational actors (DGO-Org, DGO-Clinician) supports internal governance and accountability.

### 4.2. Practical Implementation and Recommended Reporting

There is no universal standard for determining whether data is fit for use, meaning that users and context define whether data is appropriate (16). This can lead to significant discrepancies in what is considered fit for use, depending on who is conducting the assessment. Whilst this lifecycle framework does not resolve this challenge, it provides essential infrastructure for making systematic, transparent fit for use determinations, enabling data reusers to evaluate whether the quality checks performed align with their specific use case requirements.

The framework enables standardized notation for DQ reporting. The notation assists in transparency, such as; "DGO-DG-Clinician (Completeness: 94%)", distinguished from "DRO-DR-Researcher (Completeness: 87%)". This clarifies that the 7% difference reflects criteria applied by the researcher, or issues with data transformation, rather than clinician documentation gaps. For example, timeliness for a clinician in a shared care team may require instant data availability to ensure the data is fit for use, whereas researchers' use may only require monthly updates to determine whether the data is fit for use.

Application of the framework to categorize DQ checks from the case study dataset revealed that a similar DQ check may need to be conducted at different lifecycle phases or for different actors. For example, temporal plausibility refers to the believability of data. When assessed at the DG-Organization level, a policy may exist that all clinicians must report the body side of an injury; therefore, the body side is fit for reuse. Conversely, when checked at the DG-Clinician level, it is found that some clinicians never record body side, or always record the same body side, therefore, not fit for reuse. This demonstrates the framework's ability to reveal quality issues that less refined assessments would miss.

Decision rules for each data reuse case must be established for ambiguous cases. We recommend that, to ensure transparency, the reporting should be based on where the most impact on DQ was suspected to originate. For example, if the EHR system enforces the use of a code or list-based term, failures are DG-EHR issues. Yet, if clinicians have freedom to enter any code, quality is a DG-Clinician responsibility. Organizations implementing the framework should develop consensus-based guidelines appropriate to their specific EHR systems and workflows, documenting these decision rules to ensure consistent application across projects.

## 5. Limitations and future research

The framework was developed using a single dataset as a case study and will require further development relevant across healthcare settings. Some categorizations for DQ checks across the lifecycle required subjective judgment, which should be further explored across organizations. The framework provides structure for reporting where checks occurred but does not prescribe which checks should be performed, as these determinations remain context dependent.

The framework does not intend to describe every actor at each point. As health interoperability grows, more actors will be involved in patient data collection, such as carers. The framework can expand accordingly.

## 6. Conclusions

The framework enhances transparency in DQ assessments and clarifies data provenance by systematically identifying actors across lifecycle stages. It bridges the gap between comprehensive DQ assessment frameworks and practical implementation, for use by data reusers within the data-generating organisation and externally. As health data sharing expands globally, the framework provides a structure for transparent reporting of DQ characteristics, supporting improved research quality and reliable AI model development, and internal governance to enhance patient safety, clinician quality, and stakeholder reporting. The framework offers practical guidance for data reusers to better understand data fitness for reuse, and for organisations to enable targeted DQ improvement efforts across the clinical data lifecycle.

## Acknowledgements

MW acknowledges the Australian Government Research Training Program for fee support and Work Healthy Australia for providing the research dataset